\documentclass[aps,prl,floatfix,twocolumn,superscriptaddress,showpacs]{revtex4}
\usepackage{graphicx}
\usepackage{epstopdf}

\newcommand{\simlt}
      {\ifmmode       { \raisebox{-.8em}{$<$}\atop\sim}
         \else        {$\raisebox{-.8em}{$<$}\atop\sim$}
      \fi}

\begin{document}
\title{Preparation of magnetic tips for spin-polarized STM on $\mathrm{Fe}_{1+y}\mathrm{Te}$}
\author{Udai Raj Singh}
\affiliation{Max-Planck-Institut f\"ur Festk\"orperforschung, Heisenbergstr. 1, D-70569 Stuttgart, Germany}
\author{Ramakrishna Aluru}
\affiliation{Max-Planck-Institut f\"ur Festk\"orperforschung, Heisenbergstr. 1, D-70569 Stuttgart, Germany}
\author{Yong Liu}
\affiliation{Max-Planck-Institut f\"ur Festk\"orperforschung, Heisenbergstr. 1, D-70569 Stuttgart, Germany}
\author{Chengtian Lin}
\affiliation{Max-Planck-Institut f\"ur Festk\"orperforschung, Heisenbergstr. 1, D-70569 Stuttgart, Germany}
\author{Peter Wahl}
\email{wahl@st-andrews.ac.uk}
\affiliation{Max-Planck-Institut f\"ur Festk\"orperforschung, Heisenbergstr. 1, D-70569 Stuttgart, Germany}
\affiliation{SUPA, School of Physics and Astronomy, University of St. Andrews, North Haugh, St. Andrews, Fife, KY16 9SS, United Kingdom}

\date{\today}

\begin{abstract}
The interplay of electronic nematic modulations, magnetic order, superconductivity and structural distortions in strongly correlated electron materials calls for methods which allow characterizing them simultaneously - to allow establishing directly the relationship between these different phenomena. Spin-polarized STM enables studying both, electronic excitations as well as magnetic structure in the same measurement at the atomic scale. Here we demonstrate preparation of magnetic tips, both ferromagnetic and antiferromagnetic, on single crystals of FeTe. This opens up preparation of spin-polarized tips without the need for sophisticated ultra-high vacuum preparation.
\end{abstract}


\maketitle

In many unconventional superconductors, the superconducting phase is reached from a magnetically ordered state by some external tuning parameter, such as doping, pressure or chemical substitution. Superconductivity emerges in close vicinity to a magnetically ordered phase\cite{scalapino2012common}. This suggests an intimate relation between magnetism and superconductivity in these materials. Often, the phase diagrams exhibit even regimes of coexistence between the two, however the important question about whether the two coexist or compete at the microscopic level remains unresolved. One difficulty in probing their relation at the atomic scale is that most methods employed to characterize magnetic order, such as neutron scattering, probe a macroscopic sample volume, rendering statements about local phase separation difficult. A method which has been very successful to characterize both superconductivity and magnetism locally on an atomic scale is Scanning Tunneling Microscopy (STM). It has provided important information both about local variations in the superconducting properties and charge ordering in strongly correlated electron materials\cite{Lang2002, Kohsaka2007, Singh2013PRB} and, using magnetic tips in spin-polarized STM, it has also been shown to allow for characterization of magnetism at the atomic scale in nanostructures\cite{Bode2003,Wiesendanger2009}. Application of spin-polarized STM to strongly correlated materials has recently been demonstrated in the non-superconducting parent compound of the iron chalcogenide superconductors\cite{Enayat2014}, providing real space images of the magnetic structure $\mathrm{Fe}_{1+\delta}\mathrm{Te}$.
Preparing and calibrating a magnetic tip for spin-polarized STM measurements has been an important obstacle towards its application to strongly correlated electron materials.


In this work, we demonstrate preparation of spin-polarized tips and the characterization of their magnetic properties on $\mathrm{Fe}_{1+y}\mathrm{Te}$. Presence of small amounts of excess iron proves instrumental in the preparation of spin-polarized tips on this material. Specifically we show preparation of both ferromagnetic and antiferromagnetic clusters at the apex of the tip and the characterization of the magnetization of the tip-cluster as a function of field.\\

Experiments have been performed in a home-built low temperature STM operating in cryogenic vacuum at temperatures down to $1.8\mathrm K$ and in magnetic fields up to $14\mathrm T$ normal to the sample surface\cite{White2011}. Single crystals of $\mathrm{Fe}_{1+y}\mathrm{Te}$ were grown by the Bridgman method from high purity (4N) materials \cite{Liu2011}. Data has been obtained on samples with excess iron concentrations of $y=7.7\%$.
$\mathrm{Fe}_{1+y}\mathrm{Te}$ samples have been cleaved in-situ at low temperatures and then immediately inserted into the head of the STM. Measurements have been performed at a temperature of $3.8~\mathrm K$, as determined by a temperature sensor close to the STM head. After approaching the STM tip, typical STM images show a large concentration of excess iron atoms at the surface. Magnetic tips have been obtained either by picking up interstitial excess iron atoms from the surface of the material or by gentle indentation of the tip into the sample surface. The two preparation methods yield tips with predominantly ferromagnetic or antiferromagnetic behaviour.
\begin{figure}[h!]
\includegraphics[width=0.48\textwidth]{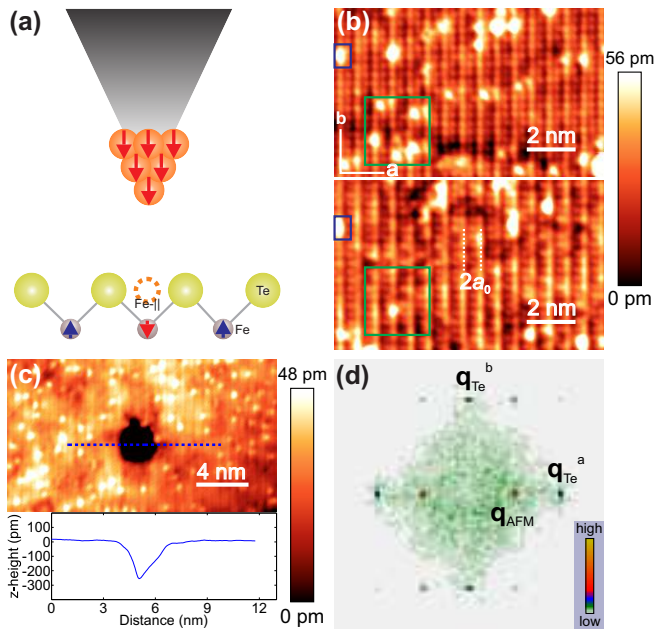}
\caption{\textbf{Preparation of Magnetic tips on FeTe.} (a) Schematic illustration of the process of picking up excess iron atoms on $\mathrm{Fe}_{1+y}\mathrm{Te}$ to prepare a ferromagnetic tip. (b) Topographic STM images obtained in the same location before (upper panel, $V_{b}=-60\mathrm{mV}$, $I_{t} = 2.0\mathrm{nA}$) and after picking up excess iron atoms ($V_{b}=-10\mathrm{mV}$, $I_{t} = 0.7\mathrm{nA}$), showing a lower concentration of excess Fe atoms (lower panel). The area marked by a green rectangle highlights excess irons atoms which were picked up. (c) STM image ($V_{b}=90\mathrm{mV}$, $I_{t} = 0.2\mathrm{nA}$) of hole left behind due to a tip indentation. The line profile across the hole is shown. The hole indicates that an $\mathrm{FeTe}$ cluster was picked up. (d) Fourier transform $\tilde{z}(\mathbf{q})$ of one of the images in b showing peaks corresponding to the square lattice of $\mathrm{Te}$ atoms ($\mathbf q^a_\mathrm{Te}$ and $\mathbf q^b_\mathrm{Te}$) and the antiferromagnetic order ($\mathbf q_\mathrm{AFM}$).}
\label{tiprep}
\end{figure}
Figure~\ref{tiprep}(a) shows a schematic illustration of the preparation of a ferromagnetic tip on the surface of $\mathrm{Fe}_{1+y}\mathrm{Te}$: by collecting excess Fe (Fe-II) atoms from the surface of the material, which are attached to the apex of the STM tip, the tip is rendered magnetic. Experiments on cobalt islands on Cu(111) show that in order to obtain a magnetic cluster which is stable at temperatures below 10K on the order of 100 atoms will be required\cite{Ouazi2012}. Fig.~\ref{tiprep}(b) and (c) show two different ways to obtain a spin-polarized tip on $\mathrm{Fe}_{1+y}\mathrm{Te}$. In Fig.~\ref{tiprep}(b), the pick-up of excess iron atoms from the surface is shown, the two images were taken before (upper panel) and after (lower panel) picking up additional interstitial excess iron atoms from the surface. The excess iron atoms appear as bright protrusions on the surface and it can be seen that there is a lower concentration of iron atoms in the lower panel of figure~\ref{tiprep}(b) indicating that some of the excess Fe atoms were picked-up by the tip. Successful preparation of a spin-polarized tip is detected by an additional modulation appearing in topographic images as seen in Fig.~\ref{tiprep}(b) with a periodicity of twice of the lattice constant of the surface tellurium atoms which corresponds to the antiferromagnetic order in the sample\cite{Enayat2014}. The second way to prepare a spin-polarized tip is by indentation into the sample surface, as shown in fig.~\ref{tiprep}(c), a process which leads to a 'hole' in the surface. Figure~\ref{tiprep}(c) shows a topographic image containing a hole which developed after a tip indentation. It clearly indicates that the tip has picked up a cluster of Fe and Te. While we have not systematically investigated which tip preparation results in specific magnetic properties of the tip, which is rendered difficult because it will depend on the history of the tip, following the above preparation recipes, we have obtained both antiferromagnetic and ferromagnetic tips. The specific magnetic properties have been characterized by measuring the field dependence of the magnetic contrast.
In Fig.~\ref{tiprep}(d), we show the Fourier transform of a topographic image obtained with a spin-polarized tip. The Fourier transform exhibits three dominant Fourier components. Two are associated with the atomic lattice at $\mathbf q^a_\mathrm{Te}$ and $\mathbf q^b_\mathrm{Te}$. The magnetic order is detected at $\mathbf q_\mathrm{AFM}=\frac{1}{2}\mathbf q^a_\mathrm{Te}$.\\

\begin{figure}[h!]
\includegraphics[width=0.47\textwidth]{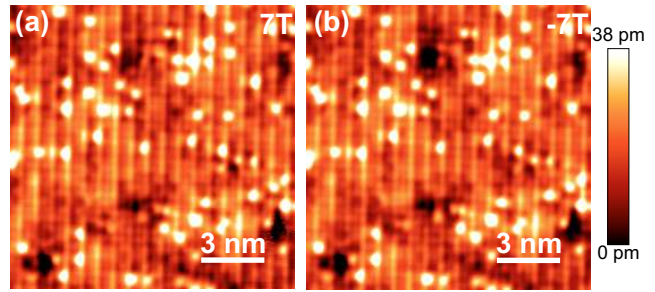}
\caption{\textbf{Magnetic field dependence of images obtained with an antiferromagnetic tip.} (a,b) STM topographies obtained at magnetic field of $+7\mathrm{T}$ (a) and $-7\mathrm{T}$ (b), both taken in the same location. The stripes due to the magnetic contrast maintain the same phase with respect to defects on the surface ($V_{b}=80\mathrm{mV}$, $I_{t}=100\mathrm{pA}$).}
\label{aftip}
\end{figure}
Fig.~\ref{aftip} shows two topographic images measured with a tip which behaves predominantly antiferromagnetic. In magnetic fields as high as $+7\mathrm T$ and $-7\mathrm T$, the phase of the magnetic contrast remains the same, and almost no change in the images is observed.\\
\begin{figure}[h!]
\includegraphics[width=0.45\textwidth]{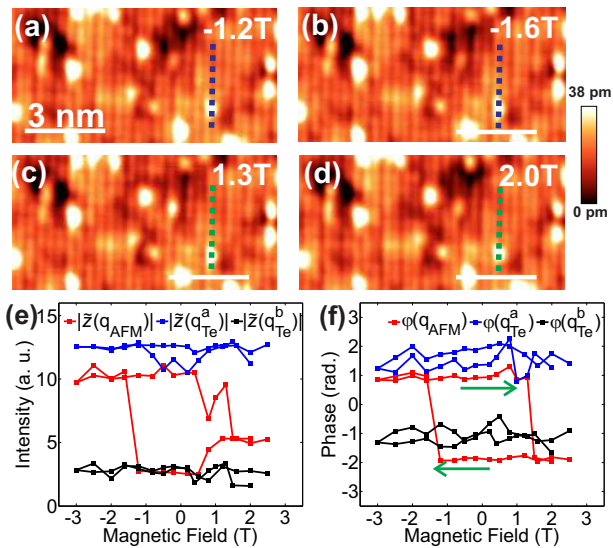}
\caption{\textbf{Magnetic field dependence of contrast obtained with a ferromagnetic tip.}~(a-d) Topographic images acquired at different magnetic fields, (a) and (b) have been obtained while ramping the field from positive to negative field and (c) and (d)  during in the opposite direction ($V_{b}=60\mathrm{mV}$, $I_{t}=200\mathrm{pA}$). The images shown are selected from a series, showing the images right before and after the tip magnetization has switched. (e) Field-dependence of the magnetic contrast, the amplitudes of the Fourier components of the magnetic modulation $|\tilde{z}(\mathbf q_\mathrm{AFM})|$ as well as for the atomic lattice vectors $|\tilde{z}(\mathbf q_\mathrm{Te}^a)|$ and $|\tilde{z}(\mathbf q_\mathrm{Te}^b)|$) are shown. The amplitude at the atomic lattice vectors remain almost constant, whereas the one at $\mathbf q_\mathrm{AFM}$ shows a change by $\sim50\%$. (f) Phase $\varphi(\mathbf{q})=\mathrm{arg}(\tilde{z}(\mathbf{q}))$ of the Fourier components at $\mathbf q_\mathrm{Te}^a$ ($=\mathbf{q}_\mathrm{CDW}$), $\mathbf q_\mathrm{Te}^b$ and $\mathbf q_\mathrm{AFM}$ as a function of field. As for the amplitude, the phase for $\mathbf q_\mathrm{Te}^a$ and $\mathbf q_\mathrm{Te}^b$ stays almost constant, while the one at $\mathbf q_\mathrm{AFM}$ reveals a hysteresis due to a ferromagnetic cluster at the apex of the tip.}
\label{fmtip}
\end{figure}
Selected images from a series of images taken with a ferromagnetic tip, with the whole series being obtained in the same location of the surface, are shown in Fig.~\ref{fmtip}(a)-(d). The series has been measured by ramping the field first from positive to negative magnetic fields (from $+2.5\mathrm T$ to $-3\mathrm T$) and then back, images have been taken in between ramping the field at fixed magnetic fields. The series of images exhibits a phase shift while ramping the field from positive to negative field and back. The images selected in Fig.~\ref{fmtip}(a) and (b) have been obtained right before (a) and after (b) the phase shift in the magnetic contrast has occurred while ramping from positive to negative fields at magnetic fields of $-1.2\mathrm T$ and $-1.6\mathrm T$. Images in panels (c) and (d) have been obtained while ramping the field back to positive fields with the stripes changing their contrast back between $1.4\mathrm T$ and $1.8\mathrm T$. To analyze the field dependence of the images in more detail, we have studied the intensity and phase of the dominant Fourier components at $\mathbf q_\mathrm{Te}^\mathrm a$, $\mathbf q_\mathrm{Te}^b$ and $\mathbf q_\mathrm{AFM}$. This analysis requires atomic registry of the images. To this end, topographic images as shown in fig.~\ref{fmtip}(a)-(d) have been aligned, especially to facilitate an analysis of the phase shift of the magnetic contrast as function of field. Both, amplitude and phase of the magnetic contrast, are expected to depend on the magnetization of the tip. Fig.~\ref{fmtip}(e) and (f) show the resulting magnetic field dependence of the amplitude and phase of the dominant Fourier components.

The amplitudes of the peaks at $\mathbf{q}_\mathrm{Te}^a$ and $\mathbf{q}_\mathrm{Te}^b$ show little magnetic field dependence (Fig.~\ref{fmtip}(e)), both stay practically constant over the complete magnetic field loop. The amplitude of the magnetic contrast at $\mathbf{q}_\mathrm{AFM}$ reveals some magnetic field dependence, it changes by about $60\%$ from its maximum value while ramping the field. The insensitivity of the amplitudes of the atomic peaks $\mathbf q_\mathrm{Te}^a$ and $\mathbf q_\mathrm{Te}^b$ to the changes in the intensity of the peak associated with the magnetic order clearly confirm that the intensity difference between the two atomic peaks is not simply an effect of higher harmonics of the modulation due to magnetic order -- but due to the charge density wave (CDW) which accompanies the magnetic order at $\mathbf q_\mathrm{CDW}=2\mathbf q_\mathrm{AFM}(=\mathbf q_\mathrm{Te}^a)$. Further, as can be seen from the Fourier components at the atomic peaks, the configuration at the apex of the tip remains stable during the measurement, except for the magnetization. If the atomic structure of the apex of the tip changed, this would be expected to influence the appearance of the atomic resolution.

The phase of the modulations associated with the atomic lattice and the CDW remains practically constant during the field sweep (Fig.~\ref{fmtip}(f)), as the amplitude, however the phase of the Fourier component of the antiferromagnetic order shows a change by $\pi$ at two magnetic fields, near $-1.6\mathrm T$ and $1.6\mathrm T$. The phase of the magnetic contrast shows clearly hysteretic behaviour of the magnetization of the tip as a function of field, as can be expected for a ferromagnetic tip. The change in the amplitude of the magnetic contrast indicates that while the magnetization of the tip reverses with the magnetic field, it does not align exactly in the opposite direction at reversed magnetic fields. Likely this is due to magnetic cluster at the apex of the tip having multiple easy magnetization axis. 

\begin{figure}[h!]
\includegraphics[width=0.47\textwidth]{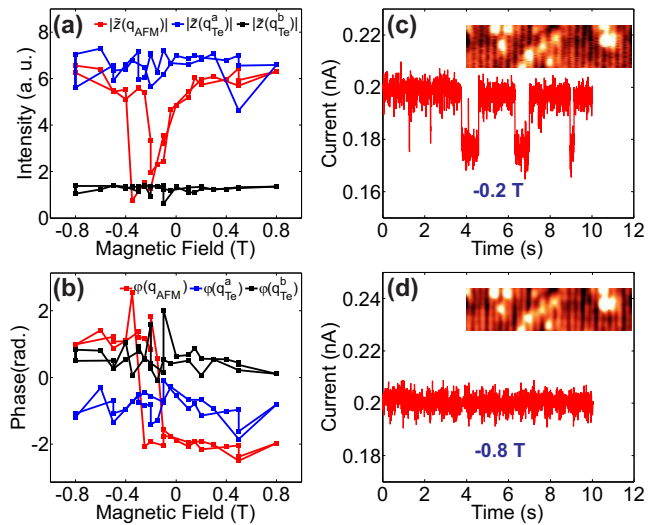}
\caption{\textbf{Magnetization dynamics of a ferromagnetic tip.}~(a) Amplitudes $|\tilde{z}(\mathbf q_\mathrm{Te}^\mathrm a)|$, $|\tilde{z}(\mathbf q_\mathrm{Te}^b)|$ and $|\tilde{z}(\mathbf q_\mathrm{AFM})|$ and (b) phase $\varphi(\mathbf q_\mathrm{Te}^\mathrm a)$, $\varphi(\mathbf q_\mathrm{Te}^b)$ and $\varphi(\mathbf q_\mathrm{AFM})$ as a function of magnetic field for a magnetic tip which shows a phase shift at fields of $-0.4\mathrm{T}$ and $-0.2\mathrm{T}$. (c) Tunneling current as a function of time measured at $-0.2\mathrm{T}$, right at the field where the phase shift occurs, obtained at $V_b=-80\mathrm{mV}$ and with open feedback loop, it shows spontaneous transitions between two magnetizations of the tip, the noise also appears in topographic images taken at the same field as shown in the inset. (c) At slightly larger or smaller fields, the current does not exhibit the transitions, shown here for a magnetic field of $-0.8\mathrm{T}$, the noise also disappears in topographic images.}
\label{timeseries}
\end{figure}
In Fig.~\ref{timeseries}(a) and (b), we present a measurement obtained with a different ferromagnetic tip showing the phase shift of the magnetic contrast at lower fields, near $-0.4\mathrm{T}$ and $+0.2\mathrm{T}$. For this tip, the intensity of the peak due to magnetic order (at $\mathbf q_\mathrm{AFM}$) is diminishing before the occurrence of the phase shift and recovers after the phase shift -- which indicates that the tip cluster has a single easy magnetization direction and its magnetization fluctuates close to the magnetic field where the phase shift occurs. The asymmetry in the magnetic fields at which the switching is observed indicates that for this tip, a ferromagnetic cluster at the apex of the tip is coupled to another magnetic cluster either with larger coercivity or which is antiferromagnetic, and hence due to exchange coupling to this second cluster the hysteresis loop becomes asymmetric. Measuring the tunnel current at magnetic fields close to the switching field reveals fluctuations of the magnetization of the tip cluster in time traces of the tunneling current. This is evidenced by jumps in the tunneling current between two states, which we attribute to switching of the magnetization direction of the tip. This is shown in Fig.~\ref{timeseries}(c), at a magnetic field just before the modulation shifts. Ramping the field to larger fields stabilizes the magnetization of the tip, Fig.~\ref{timeseries}(d) shows a measurement at $-0.8\mathrm{T}$ where no switching is observed and the current remains stable. 

\begin{figure}[h!]
\includegraphics[width=0.45\textwidth]{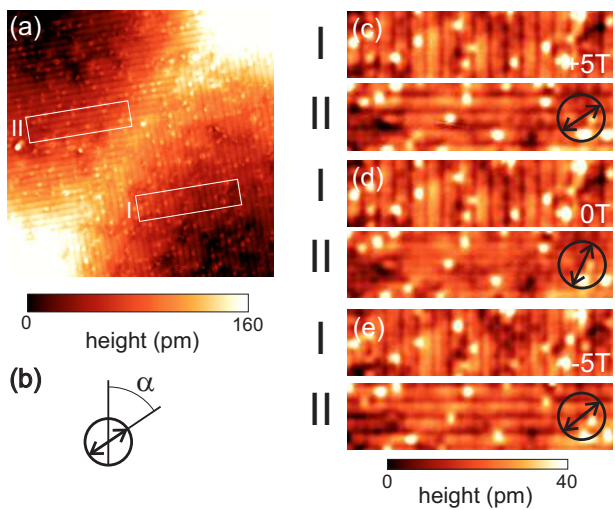}
\caption{\textbf{Characterization of In-Plane component of Magnetization} (a) STM image of a twin boundary, with the stripes on either side in orthogonal directions, (b) scheme explaining the symbol used in panels (c)-(e) for indicating the magnetization direction of the tip and the angle $\alpha$, (c)-(e) STM topographies cut out from images taken in the same location as (a) at magnetic fields of +5T, 0T, and -5T, all with the same tip ($V_{b}=80\mathrm{mV}$, $I_{t}=100\mathrm{pA}$). The in-plane magnetization direction of the tip extracted from the topographies is shown by a double arrow in the left panels (see b). Regions I and II shown in panels (c)-(e) are indicated by solid lines in (a).}
\label{inplorient}
\end{figure}
The $\mathrm{Fe}_{1+y}\mathrm{Te}$ crystals which we have characterized typically show domains of the magnetic order and monoclinic distortion, frequently domain boundaries are found where the stripes are almost orthogonal to each other in neighbouring domains. Characterization of the magnetic contrast near these twin boundaries allows determination of direction of the the in-plane component of the magnetization of the tip, because we can determine the projection onto two (almost) orthogonal directions of the magnetization. Fig.~\ref{inplorient}(a) shows a twin boundary with two domains where the magnetic order and hence the stripe pattern in the topographic image are normal to each other on the two sides of the domain boundary. If the two domains on the two sides of the boundary are denoted I and II and topographies obtained in the two $z_{\mathrm{I}}(\mathbf r)$ and $z_{\mathrm{II}}(\mathbf r)$, from $\alpha=\tan^{-1}\frac{\tilde{z}_{\mathrm{II}}\left(\mathbf{q}_\mathrm{AFM}^{\mathrm{II}}\right)}{\tilde{z}_{\mathrm{I}}\left(\mathbf{q}_\mathrm{AFM}^{\mathrm{I}}\right)}$ we can obtain the angle $\alpha$ with respect to the direction of the stripes in domain ${\mathrm{I}}$ (where $\tilde{z}(\mathbf{q})$ denotes the Fourier transform, note that $\mathbf{q}_\mathrm{AFM}^{\mathrm{I}}$ and $\mathbf{q}_\mathrm{AFM}^{\mathrm{II}}$ are almost orthogonal to each other). In fig.~\ref{inplorient}(c-e), regions cut out from topographic images obtained in the same region as the one shown in (a) but at different out of plane magnetic fields are shown. It can be seen that under a field applied normal to the surface, the magnetization of the tip rotates not only out of the plane, but also the in-plane orientation changes. Ramping the field back to zero field brings the in-plane component back to its original orientation (for this specific tip). The arrows in fig.~\ref{inplorient}(c-e) indicate the in-plane direction of the magnetization extracted as described above.

It can be observed that both in images obtained near a twin boundary as well as in the hysteresis loop, the intensity of the peak at $\mathbf{q}_\mathrm{CDW}=\mathbf{q}_\mathrm{Te}^a$ remains independent of the intensity of the peak associated with the magnetic order - confirming that the former is due to a charge modulation\cite{balatsky2010induction} rather than a higher order effect due to the magnetic order. The strong differences in the magnetic field dependence of the appearance of the stripe modulation further demonstrate that the stripe modulation is due to spin-polarized tunneling: images shown in fig.~\ref{fmtip} and \ref{aftip} have been obtained on the same sample, the differences in the magnetic field dependence are predominantly a tip property (slight tilting of the spins in the sample under magnetic fields as high as 7T can not be excluded).

In conclusion, we have shown that $\mathrm{Fe}_{1+y}\mathrm{Te}$ can be used as a material to prepare magnetic tips as well as characterize them. The availability of a preparation method for spin-polarized tips without the need for sample or tip preparation in ultra high vacuum facilitates this method to be applied in a wide variety of setups, which either only offer operation in cryogenic vacuum or lack the capability to deposit material on the tip. Both, ferromagnetic as well as antiferromagnetic tips can be prepared, allowing to study magnetic order and even metamagnetic phase transitions at high magnetic fields.

\begin{acknowledgments}
PW acknowledges funding by Max-Planck-Society and the Engineering and Physical Sciences Research Council.
\end{acknowledgments}

\label{Bibliography}

\bibliography{fete} 


%

%
%

%
%

%

\end{document}